\documentclass[amsmath,amssymb,a4paper,twocolumn,prl]{revtex4}
\usepackage{amssymb}
\usepackage{txfonts}
\usepackage{graphicx}
\usepackage{ulem}
\usepackage{color}
\usepackage{array}
\newcommand{\PreserveBackslash}[1]{\let\temp=\\#1\let\\=\temp}
\newcolumntype{C}[1]{>{\PreserveBackslash\centering}p{#1}}
\newcolumntype{R}[1]{>{\PreserveBackslash\raggedleft}p{#1}}
\newcolumntype{L}[1]{>{\PreserveBackslash\raggedright}p{#1}}

\begin{document}

\title{Pseudospins and Topological Effects of Phonons in a Kekul\'e Lattice}

\author{Yizhou \surname{Liu}$^{1,2,3}$}
\author{Chao-Sheng \surname{Lian}$^{1,2}$}
\author{Yang \surname{Li}$^{1,2}$}
\author{Yong \surname{Xu}$^{1,2,3}$}
\email{yongxu@mail.tsinghua.edu.cn}
\author{Wenhui \surname{Duan}$^{1,2,4}$}
\email{dwh@phys.tsinghua.edu.cn}

\affiliation{
$^1$State Key Laboratory of Low Dimensional Quantum Physics, Department of Physics, Tsinghua University, Beijing 100084, People's Republic of China \\
$^2$Collaborative Innovation Center of Quantum Matter, Beijing 100084, People's Republic of China \\
$^3$RIKEN Center for Emergent Matter Science (CEMS), Wako, Saitama 351-0198, Japan \\
$^4$Institute for Advanced Study, Tsinghua University, Beijing 100084, People's Republic of China}

%\date{\today}

\begin{abstract}
The search for exotic topological effects of phonons has attracted enormous interest for both fundamental science and practical applications. By studying phonons in a Kekul\'e lattice, we find a new type of pseudospin characterized by quantized Berry phases and pseudoangular momenta, which introduces various novel topological effects, including topologically protected pseudospin-polarized interface states and a phonon pseudospin Hall effect. We further demonstrate a pseudospin-contrasting optical selection rule and a pseudospin Zeeman effect, giving a complete generation-manipulation-detection paradigm of the phonon pseudospin. The pseudospin and topology-related physics revealed for phonons is general and applicable for electrons, photons and other particles.
\end{abstract}

\maketitle

One of the most exciting fields in modern condensed-matter physics is the research of topological states of quantum matter, like the quantum (anomalous or spin) Hall [Q(A or S)H] states, which has revolutionized our understanding of electronics~\cite{hasan2010, qi2011}. Topological concepts have also been applied to investigate phonon-related properties, leading to an emerging field of ``topological phononics''~\cite{prodan2009,zhang2010,wang2015prl,wang2015njp,kariyado2015,peano2015,yang2015,khanikaev2015,ni2015,liu2017prb,peng2016,fleury2016,kane2014,xiao2015,xiao2015_2,susstrunk2015,mousavi2015,pal2016,he2016,brendel2017,xia2017,nash2015,susstrunk2016,huber2016}. For phononic and acoustic systems, the QAH-like states were intensively studied, which have topologically protected one-way edge modes that can act as ideal conduction channels and diodes~\cite{prodan2009,zhang2010,wang2015prl,wang2015njp,kariyado2015,peano2015,yang2015,khanikaev2015,ni2015,liu2017prb,peng2016,fleury2016}. However, their application requires noticeably breaking time-reversal symmetry (TRS), which is still experimentally challenging despite many approaches proposed~\cite{holz1972,strohm2005,sheng2006,prodan2009,zhang2010,wang2015prl,wang2015njp,kariyado2015,peano2015,yang2015,khanikaev2015,ni2015,liu2017prb,peng2016,fleury2016}. While phonons are typically spinless~\cite{levine1962}, their properties are enriched by pseudospins. The QSH-like states can be viewed as two copies of the QAH-like states for pseudospin up and pseudospin down related to each other by a pseudo-TRS, which also have backscattering-immune edge modes~\cite{susstrunk2015,mousavi2015,pal2016}. Their realization, however, crucially relies on a dedicated design of pseudo-spin-orbit coupling (SOC). Thus, topological effects with no need of TRS breaking and pseudo-SOC are desirable for both fundamental and practical interests.

\begin{figure}
\includegraphics[width=\linewidth]{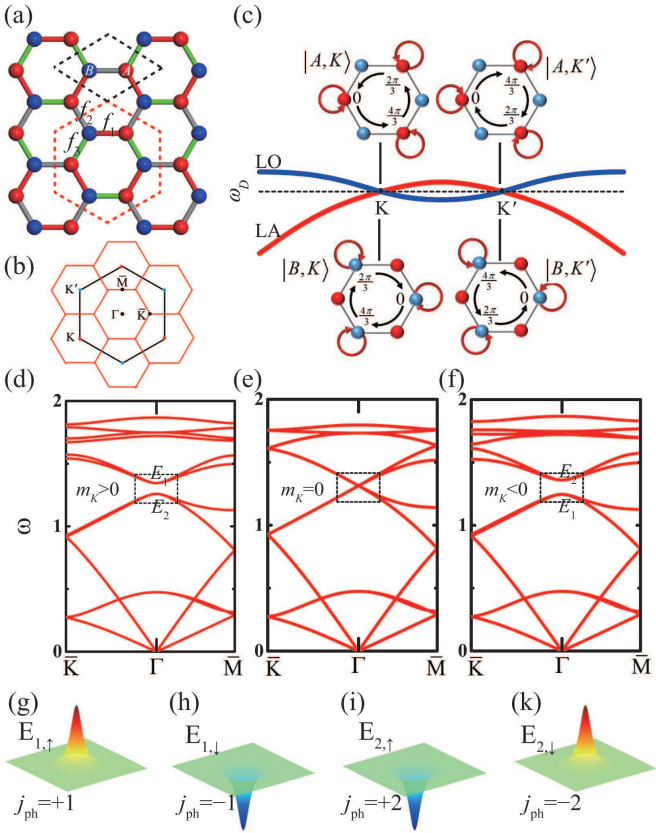}
\caption{\label{fig1}
(a) A honeycomb lattice with Kekul\'e distortion having force constants $f_{1,2,3}$ (colored red, gray, and green) with nearest neighbors, and (b) the Brillouin zone. The black and red dashed lines correspond to unit cell and $\sqrt{3}\times\sqrt{3}$ supercell, respectively. (c) Schematic of Dirac phonons formed by linear band crossings between the longitudinal acoustic (``LA'') and longitudinal optical (``LO'') branches at the $K$ and $K^\prime$ points.  The four eigenmodes ($\left| A,K \right\rangle$, $\left| B,K \right\rangle$, $\left| A,K^\prime \right\rangle$, and $\left| B,K^\prime \right\rangle$) are displayed with information about circular polarizations and phases. (d)-(f) Phonon dispersions ($\theta=\pi/4$) for $m_K=0$ and $m_K=\pm 0.05$, respectively. A band gap opens at the $\Gamma$ when $m_K \neq 0$, showing two doubly degenerate bands $E_1$ and $E_2$ whose band orders are opposite between $m_K>0$ and $m_K<0$. (g)-(k) Berry curvature distribution of $E_1$ and $E_2$ bands for pseudospin up and pseudospin down. The phonon pseudoangular momenta $j_{\textrm{ph}}$ at the $\Gamma$  are labeled.}
\end{figure}

Topological effects in solids are closely related to the Berry curvature and its integral---the Berry phase~\cite{xiao2010}. This is well demonstrated by the promising research of valleytronics~\cite{xiao2010,schaibley2016}, which discovered various topological phenomena for the valley pseudospins of electrons, like the (quantum) valley Hall effect~\cite{xiao2007,yao2008} and topological valley transport~\cite{martin2008,yao2009}. Because the topological physics is independent of particle statistics, similar extraordinary effects are applicable for phonons, introducing an intriguing field of valley phononics~\cite{zhang2015,liu2017prb,pal2017,liu2017,lu2017}. Naturally, the topological physics could be further generalized to other types of pseudospins characterized by nonzero Berry phases, pointing out new research directions. This motivates us to find novel pseudospins and topological effects for manipulating phonons in unprecedented ways.

In this Letter, we studied phonons in a honeycomb lattice with Kekul\'e distortion, namely Kekul\'e lattice, where the valley pseudospins are ill defined due to the intervalley interaction. We found a new type of pseudospin that emerges from valley-valley coupling and is characterized by a quantized Berry phase of $\pm \pi$ and a well-defined phonon pseudoangular momentum. Importantly, our study demonstrated that the phonon pseudospin not only provides a novel quantum degree of freedom to control phonons, but also leads to exotic topological effects, including topologically protected pseudospin-polarized interface states and a phonon pseudospin Hall effect. Moreover, we proposed a pseudospin-contrasting optical selection rule and a pseudospin Zeeman effect of phonons, and also predicted candidate materials to realize the Kekul\'e phonons. Our work provides a complete paradigm to generate, manipulate and detect the phonon pseudospin, which opens opportunities for future phononics.

The Kekul\'e lattice is defined as a honeycomb lattice with Kekul\'e distortion in a period of $\sqrt{3}\times\sqrt{3}$ supercell [Fig. \ref{fig1}(a)]. The corners of the original Brillouin zone (BZ) ($K$ and $K^\prime$) are folded to the center of the new BZ ($\Gamma$) [Fig. \ref{fig1}(b)]. Therefore, Dirac states at the $K$ and $K^\prime$ couple with each other, resulting in exotic physics in the distorted graphene, like chiral symmetry breaking, charge fractionalization, soliton, etc~\cite{chamon2000,hou2007,gutierrez2016}. Recently, Wu \textit{et al}. theoretically proposed the realization of QSH-like states in a Kekul\'e lattice for photons~\cite{wu2015} and electrons~\cite{wu2016,kariyado2016}. The proposal was soon generalized for acoustic systems that were examined experimentally~\cite{he2016,brendel2017,xia2017}.

To study lattice vibrations of the Kekul\'e lattice, we employed a mass-spring model that describes $A$/$B$ sublattice by masses $m_A = m_B = m$ and includes interatomic interactions with nearest and next-nearest neighbors by springs of spring constants $f_{1,2,3}=f(1+\delta_{1,2,3})$ and $f^\prime$, respectively [Fig. \ref{fig1}(a)]. For simplicity, we considered only in-plane lattice vibrations that are decoupled from out-of-plane modes in a planar structure. Let us begin with a honeycomb lattice ($\delta_{1,2,3} = 0$). The longitudinal acoustic and longitudinal optical modes form linear Dirac-like bands near the $K$ and $K^\prime$ points [Fig. \ref{fig1}(c)]. Band degeneracy at the Dirac points is protected by inversion symmetry and TRS~\cite{liu2017prb}. Phonon modes near the Dirac points, namely Dirac phonons, are described by an effective Hamiltonian $H_0$ in a basis of $p_x \pm i p_y$ orbitals~\cite{liu2017prb}:
\begin{equation}
H_0 = v_D (k_y\tau_z\sigma_x - k_x\sigma_y),
\end{equation}
where $H_0$ is referenced to the Dirac frequency $\omega_D = [(3 f + 9 f^\prime)/2m]^{1/2}$, $v_D=3fa/(8m\omega_D)$ is the group velocity ($a$ is the nearest-neighbor distance), $\textbf{k}$ is the wave vector referenced to the $K$ or $K^\prime$,  $\sigma$ and $\tau$ are the Pauli matrices with $\sigma_z=\pm1$ and $\tau_z=\pm1$ referring to sublattice $A~(B)$ and valley index $K$ ($K'$), respectively. Parameters $m=1$, $f=1$, $f^\prime = 0. 05$, and $a = 1$ were selected the same as before~\cite{liu2017prb}.

\begin{figure}
\centering
\includegraphics[width=\linewidth]{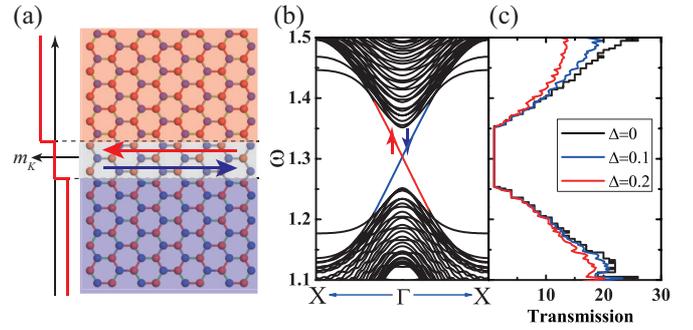}
\caption{\label{fig2}
(a) Schematic of a nanoribbon structure including an interface with opposite $m_K$ on the two sides, which has topologically protected pseudospin-polarized interface states. An armchair nanoribbon including 63 atom dimer lines with $|m_K|=0.05$ was considered, whose edge atoms in the uppermost and lowermost three lines were fixed with infinite atomic masses for eliminating trivial edge states. In the transport calculations, the transport system included a center part together with two semi-infinite contacts that were all composed of the armchair nanoribbon. Disorders were introduced over the whole center part that contains two zigzag chains (or one unit cell of the nanoribbon). (b) Phonon dispersion of the nanoribbon, showing in-gap interface states that are pseudospin-momentum locked. (c) Phonon transmission as a function of frequency $\omega$ for $\Delta = 0, 0.1, 0.2$.}
\end{figure}

Generally the Dirac point can be gapped by perturbation terms that anticommute with $H_0$, implying four types of mass terms $\sigma_z$, $\sigma_z \tau_z$, $\sigma_x \tau_x$ and $\sigma_x \tau_y$ (see Supplemental Material~\cite{suppl}). The inversion symmetry-breaking term $\sigma_z$ and the TRS-breaking term $\sigma_z \tau_z$ (i.e. the Haldane term~\cite{haldane1988}) were discussed previously~\cite{liu2017prb}. The latter two open the Dirac gap by mixing the two valleys and can be realized by the Kekul\'e distortion.

The Kekul\'e distortion introduces a perturbation term $H_K$ into the effective Hamiltonian, which is described by
\begin{equation}
H_K - \Delta_0 = \Delta_1\sigma_x\tau_x + \Delta_2\sigma_x\tau_y  =  m_K \sigma_x\tau_n,
\end{equation}
where $\Delta_0 = 2(\delta_1+\delta_2+\delta_3)\omega^\prime$, $\Delta_1 = (\delta_1 - 2\delta_2 + \delta_3)\omega^\prime$, $\Delta_2 = \sqrt{3}(\delta_3 -\delta_1)\omega^\prime$, $\omega^\prime = f/(8m\omega_D)$, $|m_K| = [\Delta_1^2 + \Delta_2^2]^{1/2}$, $\tau_n = \tau_x\cos\theta + \tau_y \sin\theta$, and $\theta$ is the azimuthal angle that is allowed to vary within $(-\pi/2,\pi/2]$ only by defining $m_K = \pm |m_K|$~\cite{suppl}. Dirac phonons in a Kekul\'e lattice, namely Kekul\'e phonons, are described by the effective Hamiltonian $H = H_0 + H_K$. Because of the band folding, a double Dirac cone is formed at the $\Gamma$ when $H_K =0$, giving a fourfold degenerate Dirac point. It is gapped into two doubly degenerate bands with dispersions $\omega_\mathbf{k} = \omega_D + \Delta_0 \pm (v^2_D|\mathbf{k}|^2+m_K^2)^{1/2}$ (labeled by $E_1$ and $E_2$) [Figs. \ref{fig1}(d)-\ref{fig1}(f)]. The band degeneracy is ensured by the $C_{3v}$ ($C_{6v}$ when $\theta = 0$) point group symmetry that has two-dimensional irreducible representations. When $\theta = 0$, $E_1$ and $E_2$ correspond to $p_x \pm i p_y$ and $d_{x^2-y^2} \pm i d_{xy}$ basis orbitals, which have odd and even parities, respectively~\cite{wu2015}. When $\theta \neq 0$, these two kinds of orbital mix with each other due to inversion symmetry breaking. $\mathbb{Z}_2$ topology of the Kekul\'e model is discussed and compared with literatures in Supplemental Material \cite{suppl}.

\textit{A new type of phonon pseudospin.}---A pseudospin operator is defined as $S=\sigma_y\tau_{n^\prime}$, where $\tau_{n^\prime} = -\tau_x\sin\theta + \tau_y\cos\theta$, satisfying the relation $[S, H] = 0$. The eigenvalue $s= \pm 1$ represents pseudospin up or pseudospin down, which is a good quantum number to label the doubly degenerate $E_1$ and $E_2$ bands. In the basis of Dirac phonon eigenmodes $\left| A,K \right\rangle$, $\left| B,K \right\rangle$, $\left| A,K^\prime \right\rangle$, and $\left| B,K^\prime \right\rangle$ [depicted in Fig. \ref{fig1}(b)], the pseudospin eigenstates at the $\Gamma$ are given by
\begin{equation}
\begin{split}
\left| E_{1,\uparrow} \right\rangle   &= \frac{1}{\sqrt{2}} (ie^{i\theta}\left| A,K^\prime \right\rangle + i \left| B,K \right\rangle), \\
\left| E_{1,\downarrow} \right\rangle &= \frac{1}{\sqrt{2}} (\left| A,K \right\rangle + e^{i\theta} \left| B,K^\prime \right\rangle), \\
\left| E_{2,\uparrow} \right\rangle   &= \frac{1}{\sqrt{2}} (\left| A,K \right\rangle - e^{i\theta} \left| B,K^\prime \right\rangle), \\
\left| E_{2,\downarrow} \right\rangle &= \frac{1}{\sqrt{2}} (-ie^{i\theta}\left| A,K^\prime\right\rangle + i \left| B,K \right\rangle).
\end{split}
\end{equation}
When $\theta \neq 0$, these pseudospin eigenstates have quantized pseudoangular momenta due to the $C_3$ rotational symmetry at the $\Gamma$~\cite{zhang2015}. The pseudoangular momentum operator, defined at $\textbf{k} = 0$, is given by $J_{\textrm{ph}} = - \sigma_z \tau_z$~\cite{suppl}, satisfying the relations $[J_{\textrm{ph}}, H] = 0$ and $[J_{\textrm{ph}}, S] = 0$. Therefore, the pseudospin eigenstates have well-defined $j_{\textrm{ph}}$ at the $\Gamma$:  $j_{\textrm{ph}} = 1$ for $\left| E_{1,\uparrow} \right\rangle$ and $\left| E_{2,\downarrow} \right\rangle$, and $j_{\textrm{ph}} = -1$ for $\left| E_{1,\downarrow} \right\rangle$ and $\left| E_{2,\uparrow} \right\rangle$. Since both $S$ and $J_{\textrm{ph}}$ anticommute with the vertical mirror plane reflections of $C_{3v}$, $s =\pm 1$ and $j_{\textrm{ph}} =\pm 1$ always appear in pairs, forming a 2D irreducible representation of the point group. When $\theta =0$, the system has a higher point group symmetry of $C_{6v}$. The above discussion remains unaffected, except that the expression of $J_{\textrm{ph}}$ becomes more complicated, and $j_{\textrm{ph}}$ of $\left| E_{2,\uparrow} \right\rangle$ and $\left| E_{2,\downarrow} \right\rangle$  changes to $+2$ and $-2$, respectively.

\begin{figure}
\centering
\includegraphics[width=\linewidth]{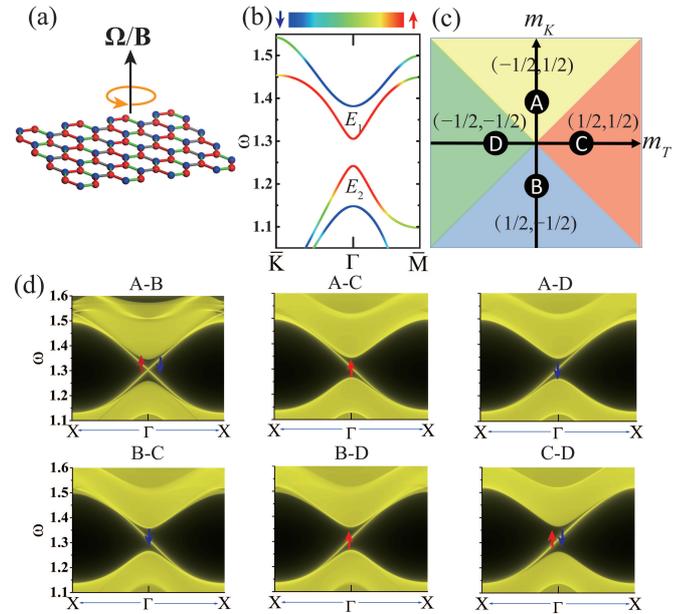}
\caption{\label{fig3}
(a) Schematic of a Kekul\'e lattice under a TRS-breaking field [e.g. Coriolis ($\pmb{\Omega}$) or magnetic ($\mathbb{B}$) field]. (b) Phonon dispersion ($m_K=0.1$) of $E_1$ and $E_2$ bands with a TRS-breaking mass of $m_T=0.05$, showing the phonon pseudospin Zeeman splitting. The pseudospin polarization was calculated by projecting wave functions onto pseudospin eigenstates. (c) A topological phase diagram for varying $m_K$ and $m_T$, which is divided into four parts characterized by different pseudospin Chern numbers ($\mathcal{C}_\uparrow, \mathcal{C}_\downarrow$). (d) Topological interface states between two distinct phases depicted in (c), calculated for interfaces between two semi-infinite regions ($|m_K|=|m_T|=0.05$) by the NEGF method.}
\end{figure}

To explore topological effects, we defined the Berry connection ${\mathbf A}_{\mathbf k} = -i \langle u_{{\mathbf k}} | \nabla_{\mathbf k} | u_{{\mathbf k}} \rangle$ and the Berry curvature ${B}_{\mathbf k} = \nabla_{\mathbf k} \times {\mathbf A}_{\mathbf k}$ for phonon eigenstates $| u_{{\mathbf k}} \rangle$. The Berry curvature distribution of Kekul\'e phonons is given for each pseudospin:
\begin{equation}\label{B_ks}
B_{{\mathbf k},s}=\pm\frac{s}{2}\frac {m_K/v_D}{[|\mathbf{k}|^2+(m_K/v_D)^2]^{3/2}},
\end{equation}
where $+$ and $-$ denote $E_1$ and $E_2$ bands, respectively. As shown in Fig. \ref{fig1}(g), $B_{{\mathbf k},s}$ is nearly zero except around the $\Gamma$, which is opposite between $E_1$ and $E_2$ and for opposite pseudospins. The integration of Berry curvature near the $\Gamma$ point gives a quantized Berry phase $\gamma_s = \pm s  \pi \textrm{sgn}(m_K)$ and thus a quantized pseudospin Chern number $\mathcal{C}_s = \pm \frac{s}{2}  \textrm{sgn}(m_K)$ when $m_K$ is small \cite{suppl}. The Berry phase gets nearly quantized for moderate $m_K$. Then it is still physically meaningful to define the pseudospin Chern number in an approximated manner \cite{suppl}. Note that the QSH-like states with $\mathbb{Z}_2 = 1$ require $\mathcal{C}_s = \pm 1$~\cite{hasan2010}, for which two copies of the Kekul\'e model are necessary.

\textit{Topologically protected pseudospin-polarized interface states.}---For an interface with opposite $m_K$ on the two sides [Fig. \ref{fig2}(a)],  the pseudospin Chern number changes sign across the interface, leading to $\Delta \mathcal{C}_s = \pm 1 $. The consequence of $\Delta \mathcal{C}_s$ is the existence of gapless phonon modes at the interface, with one pseudospin moving forward and the other backward, according to the bulk-boundary correspondence~\cite{hasan2010}. Our calculations indeed found in-gap interface states that are pseudospin-momentum locked [Fig. \ref{fig2}(b)]. This result agrees with previous data~\cite{wu2015,wu2016,kariyado2016,he2016,brendel2017,xia2017}, verifying our physical interpretation. The pseudospin-polarized interface states are ideal phonon conduction channels for topological pseudospin transport. They are robust against disorder scattering, since the conservation of pseudospin forbids backscattering. To demonstrate this feature explicitly, we introduced disorders into a ribbon structure with a topological interface and performed transport calculations by the nonequilibrium Green's function (NEGF) method~\cite{xu2008,wang2008,xu2010}. Our data show that disorders scatter most states significantly except the in-gap interface states [Fig. \ref{fig2}(c)].

Note that the topological boundary states have a tiny band gap near the $\Gamma$, because the point group symmetry that ensures the gapless feature gets slightly broken in the interface region and also partially because of the loss of strict topological protection when $m_K$ is not very small \cite{suppl}. The magnitude of the band gap ($\Delta_g$) depends on the strength of symmetry breaking. Our numerical calculations find that $\Delta_g$ can be significantly tuned by varying $\theta$, which vanishes at a critical value of $\theta_0 \approx\pm0.28\pi$ [Supplemental Fig. S5(a)~\cite{suppl}].

\textit{Phonon pseudospin Zeeman effect.}---TRS of phonons can be broken by magnetic/Coriolis fields [Fig. \ref{fig3}(a)], spin-lattice interactions in magnetic materials, etc~\cite{holz1972,strohm2005,sheng2006,prodan2009,zhang2010,wang2015prl,wang2015njp,kariyado2015,peano2015,yang2015,khanikaev2015,ni2015,liu2017prb,peng2016,fleury2016}. The influence of TRS-breaking fields on Dirac phonons is described by the Haldane term $H_T=m_T\sigma_z\tau_z$, where the coefficient $m_T$ is proportional to the field strength~\cite{liu2017prb}. Importantly, $[S, H_T] = 0$ and $[J_{\textrm{ph}}, H_T] = 0$, implying that the phonon pseudospin and its pseudoangular momentum remain well defined even when TRS is broken. The TRS-breaking field breaks the pseudospin degeneracy of $E_1$ and $E_2$ bands and introduces a pseudospin splitting, as shown in Fig.~\ref{fig3}(b). This effect is a phononic analog of the Zeeman effect, which we call the phonon pseudospin Zeeman effect. The effect can be employed to manipulate the phonon pseudospin, which is discussed more quantitatively in Supplemental Material \cite{suppl}.

One significant feature of the effect is that the pseudospin splittings of $E_1$ and $E_2$ are opposite, showing opposite pseudospin $g$ factors, which is essential to realize the QAH effect~\cite{liu2016arcmp}. Band gaps between $E_1$ and $E_2$ have different values $2(m_K-m_T)$ and $2(m_K+m_T)$ for pseudospin up and pseudospin down, respectively. The topology of this band gap is identified by $\mathcal{C}_\uparrow  = \frac{1}{2}\textrm{sgn} (m_T-m_K)$, $\mathcal{C}_\downarrow  = \frac{1}{2}\textrm{sgn} (m_T+m_K)$ and the total Chern number $\mathcal{C} = \mathcal{C}_\uparrow + \mathcal{C}_\downarrow$. As summarized in the topological phase diagram [Fig.~\ref{fig3}(c)], the competition between $m_K$ and $m_T$ gives four topologically distinct phases that are separated by critical gapless states with $|m_K|=|m_T|$. Phases $A$ and $B$ belong to the zero-Chern-number class. Phases $C$ and $D$ are not, which host the QAH-like states. Interfaces between any two of these phases all have topologically protected gapless states, comprised of one or two copies of pseudospin-polarized one-way phonon modes [Fig. \ref{fig3}(d)].

\begin{figure}
\centering
\includegraphics[width=\linewidth]{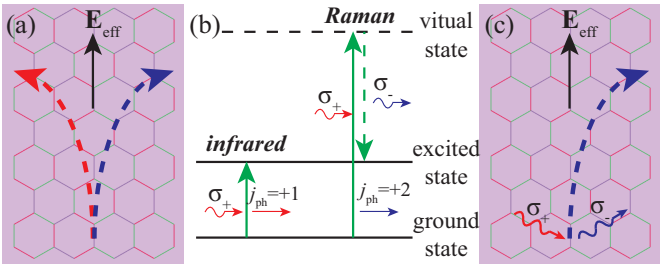}
\caption{\label{fig4}
(a) Schematic of a phonon pseudospin Hall effect under a band-edge gradient $\mathbf{E}_{\textrm{eff}}$ (black arrow). The Hall currents are opposite for pseudospin up (red curve) and pseudospin down (blue curve). (b) Schematic of pseudospin-selective phonon generation by infrared absorption and Stokes Raman scattering. Circularly polarized photons (left-handed $\sigma_-$ or right-handed $\sigma_+$) mainly excite phonons of one pseudospin that conserves the (pseudo) angular momentum. (c) By selectively generating phonons of one pseudospin (e.g. using Raman spectroscopy), a net phonon Hall current (blue curve) could be observed.}
\end{figure}

\textit{Phonon pseudospin Hall effect.}---The Berry curvature is well known as a momentum-space magnetic field, which introduces an anomalous velocity $\mathbf{v}_a \propto \mathbf{E} \times \mathbf{B}_\mathbf{k}$ transverse to the electric field, namely the anomalous Hall effect of electrons~\cite{xiao2010}. A similar effect could take place for phonons~\cite{zhang2015} in the presence of a band-edge gradient $\mathbf{E}_\textrm{eff}$ that can be generated by gradual variances of mass intensity, bonding strength, etc. The effect is opposite for the two pseudospins due to their opposite Berry curvatures, leading to a phonon pseudospin Hall effect [depicted in Fig. \ref{fig4}(a)], which could be applied to detect the phonon pseudospin.

\textit{Pseudospin-contrasting optical selection rule.}---The two phonon pseudospins have opposite pseudoangular momenta at the $\Gamma$ and, thus, can be selectively excited by circularly polarized photons (left-handed $\sigma_-$ or right-handed $\sigma_+$) due to the conservation of (pseudo)angular momentum, as demonstrated previously for the valley pseudospin~\cite{yao2008,zhang2015}. Specifically, the infrared absorption of $\sigma_+$ photons mainly excites one pseudospin with $j_\textrm{ph} = +1$, and the Stokes Raman scattering absorbing $\sigma_+$ photons mainly excites one pseudospin with $j_\textrm{ph} = +2$ by emitting $\sigma_-$ photons [illustrated in Fig. \ref{fig4}(b)]. The opposite pseudospin can be excited as well by reversing the helicity of absorbed light. This pseudospin-contrasting optical selection rule can be used to generate the phonon pseudospin, which is verified numerically by Raman intensity calculations shown in Supplemental Material \cite{suppl}. Using the pseudospin-polarized phonons as input, one can realize a net phonon Hall current [depicted in Fig. \ref{fig4}(c)], which avoids the compensation between the two pseudospins. Since the photon-phonon interactions involve only phonons near the $\Gamma$, the infrared and Raman processes are typically forbidden for Dirac phonons at the $K$ and $K^\prime$, which demonstrates the advantage of using Kekul\'e phonons.

Many materials have degenerate Dirac phonons at the $K$ and $K^\prime$, including but not limited to lattices with $C_{3v}$ or $C_{6v}$ point group symmetry (e.g. graphene, silicene, germanene, and stanene). The Kekul\'e distortion could be introduced in these materials by adatoms, substrates, external strains, etc~\cite{gutierrez2016,jin2013,zhang2014,cao2016,lee2011}. As example studies, we investigated graphene on the Sb$_2$Te$_3$ substrate and graphene under an external strain, which show the Kekul\'e distortion by forming a $\sqrt{3} \times \sqrt{3}$ reconstruction~\cite{jin2013,zhang2014,cao2016,lee2011}. Our first-principles phonon calculations found that the Kekul\'e distortion opens the phonon band gap at the $\Gamma$ as predicted theoretically (see Supplemental Fig. S4), predicting Kekul\'e phonons in realistic materials~\cite{suppl}.

In conclusion, we proposed a new type of phonon pseudospin in the Kekul\'e lattice, which offers an alternative quantum degree of freedom to control phonons and provides a platform to explore phononic topological effects independent of TRS breaking and pseudo-SOC. Based on their characteristic features of quantized Berry phases and pseudoangular momenta, we suggested some observable physical effects to manipulate, generate and detect the phonon pseudospin, including topologically protected pseudospin-polarized interface states, a phonon pseudospin Zeeman effect, a phonon pseudospin Hall effect and a pseudospin-dependent optical selection rule of phonons. The findings generalize the valley pseudospin physics that was developed originally for electrons~\cite{xiao2010,schaibley2016,xiao2007,yao2008,martin2008,yao2009} and then for phonons~\cite{zhang2015,liu2017prb,pal2017,liu2017,lu2017}. Importantly, the new phonon pseudospin is more promising than the valley pseudospin for infrared and Raman spectroscopy, which is worth further studies in the future. Furthermore, the pseudospin and topology-related physics revealed for phonons is independent of particle statistics and applicable for other particles, including electrons and photons.

\begin{acknowledgments}
Y. Liu, C.-S. Lian, Y. Li, and W. Duan acknowledge support from the Ministry of Science and Technology of China (Grant No. 2016YFA0301001) and the National Natural Science Foundation of China (Grants No. 11674188 and No. 11334006). Y. Xu acknowledges support from the National Thousand-Young-Talents Program and Tsinghua University Initiative Scientific Research Program.
\end{acknowledgments}

\end{document}